\documentstyle{article}
\begin{document}
\noindent
SAGA-HE-85-95

\noindent
June 1995

\bigskip

\bigskip

\noindent
Bulk properties of nuclear matter in the relativistic Hartree approximation
with cut-off regularization

\bigskip

\bigskip

\centerline{Kazuharu Koide, Hiroaki Kouno* and Akira Hasegawa}
\centerline{ Department of Physics, Saga University, Saga 840, Japan}

\bigskip

\centerline{and}

\bigskip

\centerline{ Masahiro Nakano}
\centerline{ University of Occupational and Environmental Health,
 Kitakyushu 807, Japan }

\bigskip

\centerline{(PACS numbers: 21.65.+f, 21.30.+y )}

\bigskip

\bigskip

\bigskip

\centerline{\bf Abstract}
A method of cut-off regularization is proposed to evaluate vacuum corrections
in nuclear matter in the framework of the Hartree approximation.
Bulk properties of nuclear matter calculated by this method are a good
agreement with results analyzed by empirical values.
The vacuum effect is quantitatively evaluated through a cut-off parameter and
its role for saturation property and compressional properties is clarified.

\bigskip

\noindent
* e-mail address: kounoh@himiko.cc.saga-u.ac.JP

\vfill\eject


In a relativistic approach for the study of nuclear matter vacuum effects
are very important corrections which can not be taken into account in any
non-relativistic formalism.
About twenty years ago, using a simple $\sigma$-$\omega$ model (Walecka model),
the mean field theory (MFT) clarified the saturation mechanism of nuclear
matter relativistically[1] and in succession vacuum effects to the saturation
property were evaluated by a method of dimensional regularization in the
framework of the Hartree approximation (RHA)[2].
It was shown that vacuum corrections made an effective nucleon mass larger and
the incompressibility of nuclear matter smaller.
A small incompressibility is desirable for experimental findings of today[3,4].
 Recently it has been reported that there is a strong correlation between an
increase of effective mass and a decrease in incompressibility[5].

Nearly ten years after a proposal of the Walecka model, corrections due to
vacuum polarization in quantum fluctuation around the mean field (the Hartree
field)  were evaluated[6].
The results were undesirable.
Corrections were larger by far than the magnitude of mean field and furthermore
 brought forth instable ghost poles (the Landou ghost) in meson propagators
which made nuclear matter unstable.
One of ad hoc but powerful recipes to escape from these disasters was to
introduce some form factors at each vertex[7].
There was another idea that the form factors should be derived from vertex
corrections[8].
These recipes are grounded on existence of internal structure of hadron.
We are afraid that vacuum corrections evaluated by the method of dimensional
regularization may be overestimated because the size of hadron is not
considered in such a regularization as the one used well in the elementary
particle physics.

When we use a conventional form factor in a calculation of polarization
insertion of vector meson, however, we need a safety device to assure the
baryon current conservation.
In this report, then, we show another recipe to estimate vacuum corrections in
the framework of Hartree approximation in simple Walecka model.
This is the first request to a method of cut-off regularization stated in the
following discussion.

A nucleon propagator $G(k)$ in the relativistic Hartree approximation
 has the following standard form
$$ G(k) = G_F(k) + G_D(k) $$
$$     = {-1 \over i\gamma_\mu k^\ast_\mu + M^\ast -i\epsilon} +
\bigr( -i\gamma_\mu k^\ast_\mu + M^\ast \bigl){i\pi\over E^\ast_k}
\theta(k_F-|\vec k|)\delta (k^\ast_0 -E^\ast_k), \eqno{(1)}$$
$$ E^\ast_k = \sqrt{ { \vec k}^2 + {M^\ast}^2 },  \eqno{(2)} $$
$$ M^\ast = M + \Sigma_S,  \eqno{(3)} $$
$$k_\mu^\ast = ( \vec k,k_4 + \Sigma_4 ) = (\vec k, ik_0 + i\Sigma_0 ),
\eqno{(4)}  $$
where the subscripts "$F$ " and "$D$" of $ G(k)$ denote the Feynman part and
the density part, respectively, $M$ and $k_F$ denote the  physical nucleon mass
and  Fermi momentum, respectively.
In the simple Walecka model the nucleon self-energies are given as follows,
$$  \Sigma_S = \Sigma_{SD} + \Sigma_{SF} = i\lambda \big({g_s\over m_s} \big)^2
\int {d^4q \over(2\pi)^4}{\rm tr}\bigr\{ G_D(q) + G_F(q) \bigl\},
 \eqno{(5)} $$
$$\Sigma_0 = i\lambda \big({g_v\over m_v}\big)^2 \int { d^4q\over(2\pi)^4}
{\rm tr} \bigr\{ \gamma_0G(q) \bigl\} = -\big({g_v\over m_v} \big)^2 \rho_B,
 \eqno{(6)} $$
where $\rho_B$ denotes the baryon density and $m_s$, $m_v$, $g_s$ and $g_v$
denote the $\sigma$-meson mass, the $\omega$-meson mass, the $\sigma$-nucleon
coupling and the $\omega$-nucleon coupling, respectively,
and $\lambda$ denotes the degeneracy, $\lambda$ = 2 for nuclear matter and
$\lambda$ = 1 for neutron matter.
The Feynman part of self-energy $\Sigma_{SF}$ is a divergent integral in the
4-dimensional momentum space while $\Sigma_{SD}$ and $\Sigma_0$ have finite
values.  In the Hartree approximation there appears another divergent integral
in the
Feynman part of baryon energy density defined as follows,
$$ \varepsilon_B = \varepsilon_{BD} + \varepsilon_{BF}
 = {\lambda\over \pi^2}\int_0^{k_F} k^2 \sqrt{k^2 + M^{\ast2} } dk -
\lambda\int{d^4k\over (2\pi)^4}{\rm tr}\bigl[ \gamma_4G_F(k) \bigr]\gamma_4.
 \eqno(7) $$

We discuss how we introduce a cut-off parameter into the two divergent
integrals in the formalism of the Hartree approximation. We start discussion
with the nucleon effective mass $M^\ast$. The effective mass is given by using
the scalar self-energy with a cut-off parameter $\Lambda$,
$$
  M^\ast
= M_0 + \Sigma_{SD}( M^\ast, \rho_B ) + \Sigma_{SF} ( M^\ast,\Lambda ) $$
$$=  M + \Sigma_{SD} ( M^\ast, \rho_B ) + \Sigma_{SF} ( M^\ast,\Lambda )   -
\Sigma_{SF}(M,\Lambda), \eqno{(8)} $$
$$  M_0 = M - \Sigma_{SF}( M, \Lambda), \eqno{(9)} $$
where $M_0 $ denotes the bare mass of nucleon.
 The last term in eq.(8) is the self-energy at zero density and is introduced
for the effective mass $M^\ast$ to be the physical mass $M$ at zero density.
 We require the bare nucleon mass $M_0$ to have cut-off dependence so that the
physical nucleon mass $M$ dose not depend on any cut-off parameter $\Lambda$.
We note that the bare nucleon mass $M_0(\Lambda)$ becomes the physical mass as
$\Lambda \rightarrow 0$.

To make progress we write the energy density as follows,
$$
\varepsilon = {1\over2} \pi^2 \alpha_V\rho_B^2 + {1\over2\pi^2}{1\over
\alpha_S}\big( M^\ast -M \big) ^2 + \varepsilon_B,
 \eqno(10)
$$
where $\alpha_S$ and $\alpha_V$ are defined,
$$
\alpha_S = \bigl({g_s\over \pi m_s} \bigr)^2, \qquad
\alpha_V = \bigl({g_v\over \pi m_v} \bigr)^2.
 \eqno(11)
$$

As the second request we require
Hugenholtz-von Hove theorem[9],
$$
 (\varepsilon + P )/\rho_B = E_N ( k_F)  \equiv E_F,
\eqno(12)
$$
where $P$ denotes pressure and $E_N(k_F)$ denotes a nucleon energy at Fermi
momentum. The condition to satisfy this identity is given by
$$
 {\partial \varepsilon \over \partial M^\ast} = 0 \Longleftrightarrow
M^\ast = M - \pi^2 \alpha_S{\partial \varepsilon_B \over \partial M^\ast}
\Longleftrightarrow  \Sigma_S = -\pi^2\alpha_S{\partial \varepsilon_B\over
\partial M^\ast} .
 \eqno(13) $$

  The relation Eq.(13) between $\Sigma_S$ and $\varepsilon_B$ is exactly
satisfied both in the density dependent part and in the Feynman part,
respectively,
 in RHA[2].
In the method of cut-off regularization, however, the same relation is not
 satisfied in Feynman part although it is satisfied in the density part.
Then we require that $\Sigma_{SF}$ is combined with $\varepsilon_{BF}$ by Eq.
(13) and obtain the expression for $\varepsilon_{BF}$ as follows,
$$
\varepsilon_{BF} = -{1\over\pi^2\alpha_S}\int\nolimits_M^{M^\ast} \Bigl[
\Sigma_{SF}(x,\Lambda) - \Sigma_{SF}(M,\Lambda) \Bigr]dx $$
$$= {\lambda \over 4}{1 \over 4\pi^2} \Bigr[ -\Lambda^4\ln\bigr\{ {\Lambda^2 +
{M^\ast}^2 \over \Lambda^2 + M^2 } \bigl\} -\Lambda^2\big({M^\ast}^2
-M^2\big) + {M^\ast}^4\ln\bigr\{ {\Lambda^2 + {M^\ast}^2\over {M^\ast}^2}
\bigl\} $$
$$ \hskip 1.8cm -M^4\ln\bigr\{ {\Lambda^2 + M^2\over M^2} \bigl\} + 4\big(
M^\ast -M \big) M\Bigr\{\Lambda^2 -M^2\ln\bigr\{ {\Lambda^2 +M^2 \over M^2
}\bigl\} \Bigl\} \Bigl],
 \eqno{(14)} $$
where
$$
\Sigma_{SF} (M^\ast,\Lambda)
= {\lambda \over 4} \alpha_S M^\ast\Bigr[\Lambda^2 - {M^\ast}^2
\ln\bigr\{ { \Lambda^2 + {M^\ast}^2 \over {M^\ast}^2} \bigl\} \Bigl],
 \eqno(15)
$$
and the lower limit $M$ in the integral is chosen so that $\varepsilon_{BF}$
 vanishes at zero density.
The scalar self-energy and the baryon energy density in the density dependent
part are given by
$$
\Sigma_{SD} = -{\lambda \over 2}\alpha_S M^\ast \Bigr[ k_FE_F^\ast - {M^\ast
}^2\ln\bigr\{ {k_F + E_F^\ast\over M^\ast} \bigl\} \Bigl],
 \eqno{(16)} $$
$$\varepsilon_{BD} = {\lambda\over8\pi^2} \Bigr[ 2k_F{E_F^\ast}^3 - {M^\ast
}^2 k_FE_F^\ast - {M^\ast}^4\ln\bigr\{ {k_F + E_F^\ast \over M^\ast} \big\}
 \Bigl], \eqno{(17)} $$
 respectively, where $E_F^\ast = \sqrt{ k_F^2 + M^{\ast2} }$ .

We make use of the saturation condition to determine the cut-off parameter at
the normal density because the cut-off parameter is independent of density.
{}From Hugenholtz-von Hove theorem the saturation condition is expressed by
$$
P =  \rho_B\bigl[ E_F^\ast + \pi^2\alpha_V\rho_B - e \bigr] = 0,
 \eqno{(18)} $$
$$ e =  M - 15.75 \quad [MeV], \eqno{(19)} $$
 at the normal density. Then we have a following relation between $M^\ast$ and
 $\alpha_V$ at the normal density.
$$
E_F^\ast + \pi^2\alpha_V\rho_B = M - 15.75 .
 \eqno(20)
$$
On the other hand, the coupling strength $\alpha_S$ is given by
$$
\alpha_S = { M^\ast - M \over \bar\Sigma_s },
\eqno(21)
$$
where $\bar\Sigma_s$ is defined as $\Sigma_s= \alpha_S\cdot\bar\Sigma_s$.
Putting these $\alpha_S$ and $\alpha_V$ into energy density equation, Eq. (10),
 we have a kind of relation between the effective mass $M^\ast$ and the cut-off
 parameter $\Lambda$ which satisfies the saturation condition at the normal
density.

In Fig. 1, we show the saturation curves for several sets of $M^\ast$ and
$\Lambda$. The other bulk properties of nuclear matter can be calculated for
each set and are summarized in Table.

\centerline{ Fig. 1 and Table}

First of all we discuss the magnitude of cut-off parameter $\Lambda$ which
looks very small at first glance.
We try to calculate $\Sigma_{SF}$ again by introducing the following
conventional monopole type of form factor into vertex,
$$ F_{NN\sigma}(p,q) = {\Lambda_N^2 - M^2\over p^2 + \Lambda_N^2 }\cdot
{\Lambda_N^2 -M^2 \over (p + q)^2 + \Lambda_N^2 }\cdot{\Lambda_\sigma^2 \over
q^2 +\Lambda_\sigma^2 },
 \eqno(22) $$

\centerline{Fig. 2}

\noindent
We note that we use this form factor for $\Sigma_{SF}(M,\Lambda$) at zero
density while we modify it for $\Sigma_{SF}(M^\ast,\Lambda)$  at finite density
 by replacing $M$ with $M^\ast$ in Eq.(22).
We make a comparison between the new result and $\Sigma_{SF}(M^\ast,\Lambda)$,
 and obtain the following relation between two kind of cut-off parameter,
$$ \Lambda_N^2 = M^2 + \Lambda^2 , \qquad \bigl( {\Lambda_N^\ast}^2 ={M^\ast}^2
 +  \Lambda^2 \bigr),
 \eqno(23) $$
where another cut-off parameter $\Lambda_\sigma$ dose not participate in
Hartree calculation because $q^2 = 0$.
When we take about a half of nucleon mass as the cut-off parameter $\Lambda$,
for example, we have familiar values for the cut-off $\Lambda_N$ of the
monopole type of form factor.

Next, we are very interested in the saturation curves in Fig. 1.
The saturation curves are softer as an increase of $\Lambda$.
Starting from the most stiff curve (MFT) obtained without vacuum effect
($\Lambda=0$), there exist the curves with smaller values of incompressibility
less than 200 MeV.
We can understand the saturation mechanism as a stable balance of three force,
i.e., the attractive $\sigma$-meson, the repulsive $\omega$-meson and another
repulsive vacuum effect.
Contributions of meson to energy density depend strongly on the baryon density
whereas contributions of vacuum effect depend weakly on the baryon density.
So, since the energy density is fixed at the normal density, if the vacuum
contribution to the energy density is larger, the $\omega$-meson contribution
is smaller.
Thus, the increase of $\Lambda$ dulls the density dependence of energy density
in the neighbourhood of the normal density.
 This is the reason that the incompressibility $K$ becomes small if $\Lambda$
increase, as shown in Table.

Also we make a remark on the relation between $M^\ast$ and $\Lambda$.
The effective mass depends on the attractive $\sigma$-nucleon coupling $g_s$
and the repulsive vacuum parameter $\Lambda$.
The former makes $M^\ast$ small and the latter dose $M^\ast$ larger. The
effective mass continues to increase if $0 < \Lambda/M < 0.4$, reaches to the
maximum
at $\Lambda/M\sim 0.4$ and decreases if $0.5 < \Lambda/M < 0.56$.
Then, the vacuum effect for the effective mass is largest at $\Lambda/M \sim
0.4$. It is reasonable that the repulsive $\omega$-nucleon coupling strength
$g_v$ shows a minimum value at $\Lambda/M \sim 0.4$ ( the maximum effect of
vacuum ). It can be observed through the effective mass that the parameter
$\Lambda$ and $g_v$ are complement each other as the two repulsive effects.

The effective mass $M^\ast$ decreases as the incompressibility $K$ decreases
less than 400 MeV.
The small $M^\ast$ is rather desirable since the empirical spin-orbit
splitting in light nuclei supports $M^\ast = 0.6M$ [10].
The skewness $K^\prime$ ( the third order derivative of saturation curve at
the normal density in Ref. [5]) can also be calculated.
Using this $K^\prime$, the Coulomb coefficient $K_c$ is given by
$$
K_c = -{3q_e^2\over5r_0} \Bigl({9K^\prime\over K} + 8\Bigr) , \qquad
r_0 = \bigl( {3\over 4\pi \rho_0 } \bigr)^{1/3},
 \eqno(24)
$$
based on the scaling model[11], where $g_e$ denotes the proton electric
charge. $K_c$ is the coefficient of leptdermous expansion[11],
$$
K(A,Z) = K + K_{sf}A^{-1/3} + K_{vs}I^2 + K_cZ^2A^{-4/3} + \cdots,
 \qquad I = 1-2{Z\over A},
\eqno(25)
$$
where $K_{sf}$ and $K_{vs}$ are the surface-term coefficient and the
volume-symmetry-term coefficient, respectively.
These coefficients are determined from the giant monopole resonance(GMR) data
of many nuclei.
In fig. 3, we show the $K - K_c$ relation together with results  analyzed by
empirical values in Table 3 of [3] and in Table IV of [4].
We have a fine agreement with results analyzed empirically if 200 MeV $< K <$
350 MeV.

\centerline{Fig. 3}

The symmetry energy $a_4$ in Table includes the $\rho$-meson contribution which
depends on the square of the ratio of coupling strength to mass of $\rho$-meson
and makes $a_4$ increase to about 30 MeV, i.e., without this contribution
$a_4$ becomes 20.6 MeV at $K$ = 300 MeV, in Table.

The $K- K_{vs}$ relation is shown in Fig. 4. The quantity $K_{vs} $ is given by
$$
K_{vs} = K_{sym} - L \Bigl( 9{ K^\prime\over K} + 6 \Bigr),
 \eqno(26)
$$
in the scaling-model, where $L$ and $K_{sym}$ are the first and second order
derivative of asymmetry energy, respectively ( see the detailed definitions
in Ref. [5]). Also we have another fine agreement with empirical values[3,4]
if 250 MeV $< K <$ 400 MeV.

In summary, we proposed the method of cut-off regularization to evaluate
the vacuum corrections in nuclear matter in the framework of the Hartree
approximation.
We found that this method, RHAC( Relativistic Hartree Approximation with
Cut-off regularization), can prepare the values from 200 MeV to 546 MeV for
nuclear
incompressibility in spite of a few adjustable parameters.
So we note that the RHAC method is a very useful phenomenological one under the
 present situation that there is much uncertainty in the experimental
determination of compressional properties.

We made quantitative analysis of the vacuum correction by the cut-off
parameter. The results are summarized as follows.

(1) An increase of $\Lambda$ means an increase of vacuum correction.
The parameter $\Lambda$ can be connected with the cut-off parameter of the
conventional monopole type of form factor.

(2) The vacuum correction gives the repulsive effect both to the effective
nucleon mass and to the baryon energy density.
The saturation property is yielded by the interplay among the attractive
$\sigma$-meson, the repulsive $\omega$-meson and the repulsive vacuum effect.
The repulsive vacuum effect makes the nucleon incompressibility small
because of its weak dependence on the baryon density.

(3) The calculated asymmetry energies in Table agree well with the empirical
values if $C_\rho^2 =(g_\rho M/m_\rho )^2$ = 54.71 is used as the
$\rho$-meson coupling strength[5].

(4)  The calculated curve on the $K - K_c$ plane is a fine agreement with the
empirical candidates in region 200 MeV $< K <$ 350 MeV and also the curve on
the $K - K_{vs}$ plane is a good agreement with the empirical candidates in
region  250 MeV $< K <$ 400 MeV.
Therefore, to account for $K_c, K_{vs}$ and $a_4$ simultaneously, the RHAC
method is valid if 250 MeV $< K <$ 350 MeV.

\centerline{\bf Acknowledgment}
The authors would like to acknowledge instructive discussions with Dr. T.
Harada.
They would like to thank T. Mitsumori and N. Noda for useful discussions.
The authors are grateful to the members of nuclear theorist group in Kyushyu
 district in Japan for continuous encouragement.

\centerline{\bf References }

\noindent
[1]. J. D. Walecka, {\it Ann. of Phys.} {\bf 83}, 491(1974).\par
\noindent
[2]. S. A. Chin, {\it Ann. of Phys.} {\bf 108}, 301(1977).\par
\noindent
[3]. J. M. Pearson, {\it Phys. Lett.} {\bf B271}, 12(1991).\par
\noindent
[4]. S. Shlomo and D. H. Youngblood, {\it Phys. Rev.} {\bf C47}, 529(1993).
\par
\noindent
[5]. H. Kouno, N. Kakuta, N. Noda, K. Koide, T. Mitsumori, A. Hasegawa  \par
and M. Nakano, {\it Phys. Rev.} {\bf C51}, 1754(1995), \par
H. Kouno, K. Koide, T. Mitsumori, N. Noda, A. Hasegawa and M. Nakano, \par
 to be published in {\it Phys. Rev.} {\bf C}. \par
\noindent
[6]. R. J. Furnstahl, R. J. Perry and B. D. Serot, {\it Phys. Rev.} {\bf C40},
321(1989),\par
 K. Lim, Ph.D. Thesis in the Dept. of Phys., Indiana Univ., (1990) ,\par
 K. Lim and C. J. Horowitz, {\it Nucl. Phys.} {\bf A501}, 729(1989).\par
\noindent
[7]. M. Prakash, P. J. Ellis and J. I. Kapusta, {\it Phys. Rev.} {\bf C45},
2518(1992).\par
 V. K. Mishra, G. Fai, P. C. Tandy and M. R. Frank, {\it Phys. Rev.} {\bf C46},
1143(1992),
\par
 J. A. McNeil, C. E. Price and J. R. Shepard, {\it Phys. Rev.} {\bf C47},
1534(1993).\par
\noindent
[8]. M. P. Allendes and B. D. Serot, {\it Phys. Rev. } {\bf C45},
2975(1992),\par
B. D. Serot and H.-B. Tang, {\it Phys. Rev.} {\bf C51}, 969(1995).\par
\noindent
[9]. N. M. Hugenholtz and L. van Hove, {\it Physica} {\bf 24}, 363(1958). \par
\noindent
[10]. A. R. Bodmer and C. E. Price, {\it Nucl. Phys. } {\bf A505},
123(1989).\par\noindent
[11]. J. P. Blaizot, {\it Phys. Rep.} {\bf 64}, 171(1980).\par

\vfill\eject

\centerline{ \bf Table and Figure Captions }

\noindent
Table \quad The numerical results for parameter sets of ($M^\ast$, $\Lambda$).
$K$, $K^\prime$, $K_c$, $a_4$, $K_{vs}$, $K_{sym}$ and $L$ are shown in MeV.
The $a_4$, $K_{vs}$, $K_{sym}$ and $L$ depend on value of $C_\rho = g_\rho
M/m_\rho$ where $C_\rho^2$ = 54.71[5].

\noindent
Fig.1 \quad The $k_F$-dependence of binding energy. All curves take a minimum
value( -15.75 MeV ) at the normal Fermi momentum ( 1.35 fm$^{-1}$ ). The bold
solid line, the bold dash-dotted line, the bold dashed line, the bold dotted
line, the  dash-dotted line, the dashed line, the dotted line and the solid
line are the results for $K$ = 546, 500, 450, 400, 350, 300, 250 and 200  MeV
in Table , respectively. The bold solid line is the result of MFT without the
vacuum effect.

\noindent
Fig. 2 \quad Diagrammatic representation of the form factor
at $NN\sigma$ vertex.
The solid ( wavy ) lines denote nucleon ( $\sigma$-meson ) propagators.
The form factor is shown as the circle at vertex.

\noindent
Fig.3 \quad The $K - K_c$ relation.
The crosses with error bars are results in ref. [3] and the solid squares  are
 the data from the table IV in ref. [4].

\noindent
Fig.4 \quad The $K - K_{vs}$ relation.
The crosses with error bars are results in ref. [3] and the solid squares  are
 the data from the table IV in ref. [4].

\vfill\eject

\large

\hspace*{-2.8cm}
  \begin{tabular}{lrrrrrrrrrr}
                                    \hline
   \ $\Lambda / M$ & $K$ & $K^\prime$ & $K_c$  & $a_4$ & $K_{vs}$
      & $K_{sym}$ & $L$~~ & $M^\ast / M$ & $g_s$ & $g_v$ \\ \hline
   \ 0.0000 & 546 & 226.39 & -8.971 & 30.2 & -890.4 &  86.3 & 100.36
            & 0.5470 & 10.438 & 12.899 \\
   \ 0.2827 & 500 & 160.77 & -8.330 & 29.5 & -775.3 &  72.0 & 95.27
            & 0.5722 & 10.447 & 12.498 \\
   \ 0.3884 & 450 & 101.96 & -7.676 & 29.0 & -671.5 &  65.6 & 91.69
            & 0.5924 & 10.779 & 12.165 \\
   \ 0.4892 & 400 &  54.71 & -7.058 & 29.0 & -587.6 &  79.8 & 92.29
            & 0.5919 & 11.903 & 12.173 \\
   \ 0.5407 & 350 &   0.05 & -6.118 & 29.7 & -463.3 & 127.9 & 98.50
            & 0.5650 & 13.467 & 12.615 \\
   \ 0.5559 & 300 & -78.33 & -4.320 & 30.5 & -197.4 & 186.0 & 105.02
            & 0.5398 & 14.535 & 13.011 \\
   \ 0.5609 & 250 & -177.10 & -1.242 & 31.1 & 288.0 & 246.4 & 110.74
            & 0.5202 &  15.262 & 13.308 \\
   \ 0.5624 & 200 & -293.20 & 3.972 & 31.6 & 1142.4 & 308.9 & 115.86
            & 0.5044 &  15.801 & 13.542 \\ \hline

  \end{tabular}
\begin{center}
  Table
\end{center}
%
%

\end{document}